\documentclass[final,3p,times]{elsarticle}
\usepackage{amssymb}
\usepackage{amsthm}
\usepackage{amsmath}
\usepackage{xfrac}
\usepackage{subcaption}

\usepackage[colorlinks]{hyperref}
\usepackage{cancel}
\usepackage{xcolor}
\usepackage{float} % needed if you want [H] later
\usepackage[figurename=Fig.,labelfont=bf]{caption}

\usepackage[ruled,vlined]{algorithm2e} %to write algorithms
\makeatletter
\newcommand*{\algotitle}[2]{%
	\stepcounter{algocf}%
	\hypertarget{algocf.title.\theHalgocf}{}%
	\NR@gettitle{#1}%
	\label{#2}%
	\addtocounter{algocf}{-1}%
}
\makeatother
\usepackage{framed}
\usepackage{multicol}
\usepackage{cleveref}
\usepackage{hyperref}
\usepackage{longtable}
\usepackage[dvipsnames]{xcolor}
\usepackage{comment}
\usepackage{array}
\usepackage{multirow}
\usepackage{booktabs}
\usepackage{soul}
\usepackage{tabularx}

\immediate\write18{%
	makeindex -s nomencl.ist -o \jobname.nls -t \jobname.nlg \jobname.nlo%
}

\usepackage{nomencl}
\makenomenclature

\begin{document}
 
	\begin{frontmatter}
	
        \title{An efficient open-source framework for high-fidelity 3D surface topography and roughness prediction in milling}
		
        \author[add1]{Hadi Bakhshan\fnref{fn1}}
        \author[add1,add2]{Sima Farshbaf\fnref{fn2}}
        \author[add3,add4]{Adrián Travieso-Disotuar}
        \author[add5]{Luciano Mijaíl Villarreal}
        \author[add1]{Fernando Rastellini Canela}
        \author[add1,add6]{Josep Maria Carbonell}
        
        \fntext[fn1]{Email: hbakhshan@cimne.upc.edu}
        \fntext[fn2]{Email: sima.farshbaf@upc.edu}
        
		\address[add1]{Centre Internacional de Mètodes Numèrics a l'Enginyeria (CIMNE), Campus Norte UPC, 08034 Barcelona, Spain}
        \address[add2]{Universitat Politècnica de Catalunya (UPC), Campus Norte UPC, 08034 Barcelona, Spain}
        \address[add3]{Eurecat, Centre Tecnològic de Catalunya, Unit of Metallic and Ceramic Materials, Plaça de la Ciència, Manresa, Barcelona, 08243, Spain}
        \address[add4]{Universitat Politècnica de Catalunya, Escola d'Enginyeria de Barcelona Est, Av. d'Eduard Maristany, 10-16, Barcelona, 08019, Spain}
        \address[add5]{Grupo Sevilla Control, Aerospace Engineering, Machining and Assemblies, R+D+I Department, 41007 Sevilla, Spain}
 		\address[add6]{Mechatronics and Modelling Applied on Technology of Materials (MECAMAT) group. Universitat de Vic-Universitat Central de Catalunya (UVic-UCC), C. de la Laura 13, 08500 Vic, Spain} 

%---------------------------------------------------------------------------------------------
%---------------------------------------------------------------------------------------------
%---------------------------------------------------------------------------------------------

\begin{abstract}

    With the emergence of data-driven approaches in science, there is growing interest in their application to manufacturing, particularly in surface precision engineering. However, generating large datasets required for model training is often impractical experimentally due to high costs and the time-intensive nature of measurements. High-fidelity synthetic datasets offer a viable alternative if they can be generated both efficiently and accurately. To address this challenge, this paper presents an efficient framework for generating accurate 3D surface topographies and roughness indicators in milling operations using numerical methods. First, a conventional topography prediction model is developed based on the forward solution method (FSM). Building on this, an optimized computational algorithm is proposed to establish an efficient FSM with significantly improved performance. The model is validated against two independent sets of experimental results, assessing both prediction accuracy and computational efficiency. The results demonstrate acceptable prediction errors and an average computational speedup of 42.2×. The proposed open-source model provides a generalizable framework for large-scale analysis, enabling the generation of extensive datasets for data-driven surrogate modeling.

\end{abstract}

%---------------------------------------------------------------------------------------------
%---------------------------------------------------------------------------------------------
%---------------------------------------------------------------------------------------------

		
	\end{frontmatter}
	
	%% \linenumbers

%---------------------------------------------------------------------------------------------
%---------------------------------------------------------------------------------------------
%---------------------------------------------------------------------------------------------

	%% main text
\section{Introduction}
	\label{introduction}

    The rapid advancements of industries such as aerospace \cite{liu2023research}, energy systems \cite{abellan2024review}, optics \cite{ciambriello2022influence}, and semiconductor manufacturing \cite{song2022review, gao2025predictive} necessitate increasingly stringent requirements for the surface quality of critical components. Many high-performance parts operate in extreme environmental conditions characterized by high temperatures, moisture, and corrosive environments, which highlights the significant role of the surface integrity of the machined surface in determining component reliability and operational lifespan \cite{yang2023temperature, xiao2023review, chen20253d}. Surface topography, which describes the three-dimensional characteristics of a machined surface, directly influences key functional properties such as friction \cite{berglund2010milled}, contact behavior \cite{mu2019feasibility}, wear resistance \cite{ghosh2015novel, bhushan2022effect, chang2024comprehensive}, and fatigue performance \cite{pegues2018surface, sanaei2020analysis}. Therefore, surface roughness and related topographical features are commonly employed as important indicators of machining quality.

    The milling process is a widely used subtractive manufacturing method due to its versatility, productivity, and economic advantages \cite{sun2023review}. However, achieving a target surface quality in milling typically requires numerous experimental trials, which can be both time-consuming and costly \cite{chen20253d}. Consequently, reliable predictive approaches for surface topography and roughness are essential to improve process understanding and enable more efficient selection of machining parameters \cite{zhao2023machined, yao2024extreme}. Several methods have been proposed to predict surface roughness and topography, which can be broadly categorized into four main groups.

    First, empirical methods aim to establish statistical relationships between machining parameters and surface quality indicators through regression analysis based on experimentally measured data. Various linear, polynomial, or power-law relationships are formulated to correlate surface roughness parameters with machining variables such as cutting conditions, tool parameters, workpiece properties, and environmental factors \cite{he2018influencing}. However, approaches that rely heavily on experimental observations are often limited in generality and dependent on significant resources. This makes them inefficient for covering a wide range of process conditions, especially considering the large number of influential parameters involved in machining processes.

    Second, analytical and mechanistic approaches describe surface generation based on the kinematics of the cutting tool, tool geometry, and machining parameters. These models capture the relative motion between the cutter and the workpiece, where the machined surface is considered as the envelope of the tool trajectory, resulting in the generation of surface topography \cite{sun2023review, arizmendi2019modelling, yang2015surface}. Analytical formulations are often used to derive simplified expressions for common roughness parameters, such as average areal or line roughness, by approximating the surface profile generated by the periodic engagement of the cutting edges.

    Third, data-driven methods rely on large datasets and are widely employed for surface roughness prediction \cite{yang2024review}, while their application to full surface topography prediction remains relatively limited. These approaches typically learn a functional mapping between machining parameters such as cutting speed, feed rate, and depth of cut and surface roughness values using machine learning (ML) or deep learning (DL) techniques. Representative models include artificial neural networks (ANNs) \cite{zain2010prediction, boga2021proper}, support vector machines (SVMs) \cite{yeganefar2019use, abu2017surface}, and other learning-based algorithms. In such cases, both input and output data are generally numerical. For image-based data, techniques such as convolutional neural networks (CNNs) \cite{rifai2020evaluation, chen2021visual}, recurrent neural networks (RNNs) \cite{shang2024spatiotemporal, wang2022online}, and generative models \cite{wang2022novel} have been employed to extract features from surface topography images. These models can process two-dimensional pixel-based images or three-dimensional voxel-based representations of surface morphology, and in some cases incorporate machining parameters as additional inputs to predict surface characteristics. However, data-driven methods typically require large datasets for training, and generating sufficient experimental data is often impractical. Consequently, synthetic datasets generated from analytical or numerical approaches are frequently used to support the development of such models.

    Fourth, numerical approaches provide an effective framework for modeling machined surface topography when analytical formulations become insufficient to capture the complex interactions occurring during the cutting process. In general, these methods discretize the spatial and temporal domains into finite intervals, allowing the surface generation process to be reconstructed step by step. Such approaches enable the incorporation of additional factors in the machining process, including tool run-out, tool geometry, multi-tooth engagement, and complex tool kinematics, which are difficult to represent in purely analytical formulations. Several studies have attempted to classify numerical methods for machined surface topography prediction \cite{sun2023review, wang2021modified, xu20203d, zhang2008new}. A commonly adopted classification categorizes numerical approaches according to their computational flow into forward solution methods (FSM), backward solution methods, and forward–backward solution methods, as suggested by \cite{wang2023high}.
    
    Forward methods track the cutting-edge trajectory over discretized time steps and generate the surface by selecting the minimum height at each workpiece point. These methods offer straightforward implementation but often involve high computational cost due to the fine temporal discretization required \cite{zhou2018surface, li2013surface}. Backward methods, in contrast, start from discretized workpiece points and solve for the corresponding tool position and cutting time. This approach reduces the need for trajectory discretization but typically requires iterative solutions of nonlinear or transcendental equations \cite{arizmendi2019modelling, liu2020online, gao2006numerical}. Hybrid forward–backward methods combine both strategies by representing the swept surface of the cutting-edge and computing intersections with surface points. In doing so, they partially reduce computational effort while maintaining geometric accuracy \cite{torta2020surface, shujuan2019geometrical}. Despite these advancements, existing approaches still face challenges in balancing surface resolution, computational efficiency, and accurate correspondence between tool motion and generated surface points.

    Surface topography models are generally evaluated based on two key aspects: prediction accuracy and computational efficiency. Research on model accuracy is relatively mature, as many existing approaches achieve comparable levels of accuracy under similar machining conditions \cite{wang2023high}. In contrast, computational efficiency has received comparatively less attention, particularly for large-scale milling simulations where high-resolution surfaces must be generated for cutters with multiple inserts (indexable tools) \cite{muthuswamy2021experimental, pimenov2019effect} and for simulations involving large workpiece areas \cite{liu2019coupled}. Furthermore, with the emergence of data-driven approaches, the demand for high-fidelity and computationally efficient surface simulations has increased. Efficient simulation models enable systematic exploration of machining parameters and facilitate the generation of extensive datasets of surface topographies and associated process conditions, which are essential for training predictive algorithms and supporting optimization in modern data-driven manufacturing.
    
    To address limitations in existing research, this paper presents an efficient open-source framework based on the FSM for predicting the three-dimensional surface topography of face-milled parts and their associated roughness metrics. To overcome the computational challenges of high-resolution simulations, the proposed efficient FSM (EFSM) is designed to maximize computational performance while maintaining usability. The computational core is implemented in a compiled programming language with parallelization techniques and is bound to a scripting interface to enable flexible user interaction. The EFSM components including kinematic equations, discretization, and tool-motion tracking are implemented with preallocated data structures to reduce computational overhead. This modular architecture separates the intensive computation from interface and control tasks, enabling efficient generation of high-fidelity surfaces and large datasets for surface analysis and data-driven manufacturing applications.

%---------------------------------------------------------------------------------------------
%---------------------------------------------------------------------------------------------
%---------------------------------------------------------------------------------------------

\section{Conventional surface topography model}
    \label{sec_conventional_model}

    This section describes the FSM as a conventional methodology to predict the surface topography of a milling process, specifically face milling in this study. The process of generating and predicting surface topography using FSMs aims to simulate the physical tool-workpiece interaction based on the kinematic equations of the tool path. It calculates the progressive removal of material as the cutting-edge sweeps across the workpiece, and the remaining envelope defines the resulting texture or topography.
    
    In general, FSM involves three underlying mechanisms: acquiring the kinematic equations of the cutting-edge movements, applying a discretization procedure, and following the trace of the cutting-edge to calculate its trajectory path. This section delves into these steps based on the methodology proposed in \cite{zhang2018modeling, wang2023high}.

    \subsection{Describing cutting-edge motion}
	    \label{subsec_describing_cutting-edge_motion}

\begin{figure}[ht]%
            \centering
            \includegraphics[width=0.99\textwidth]{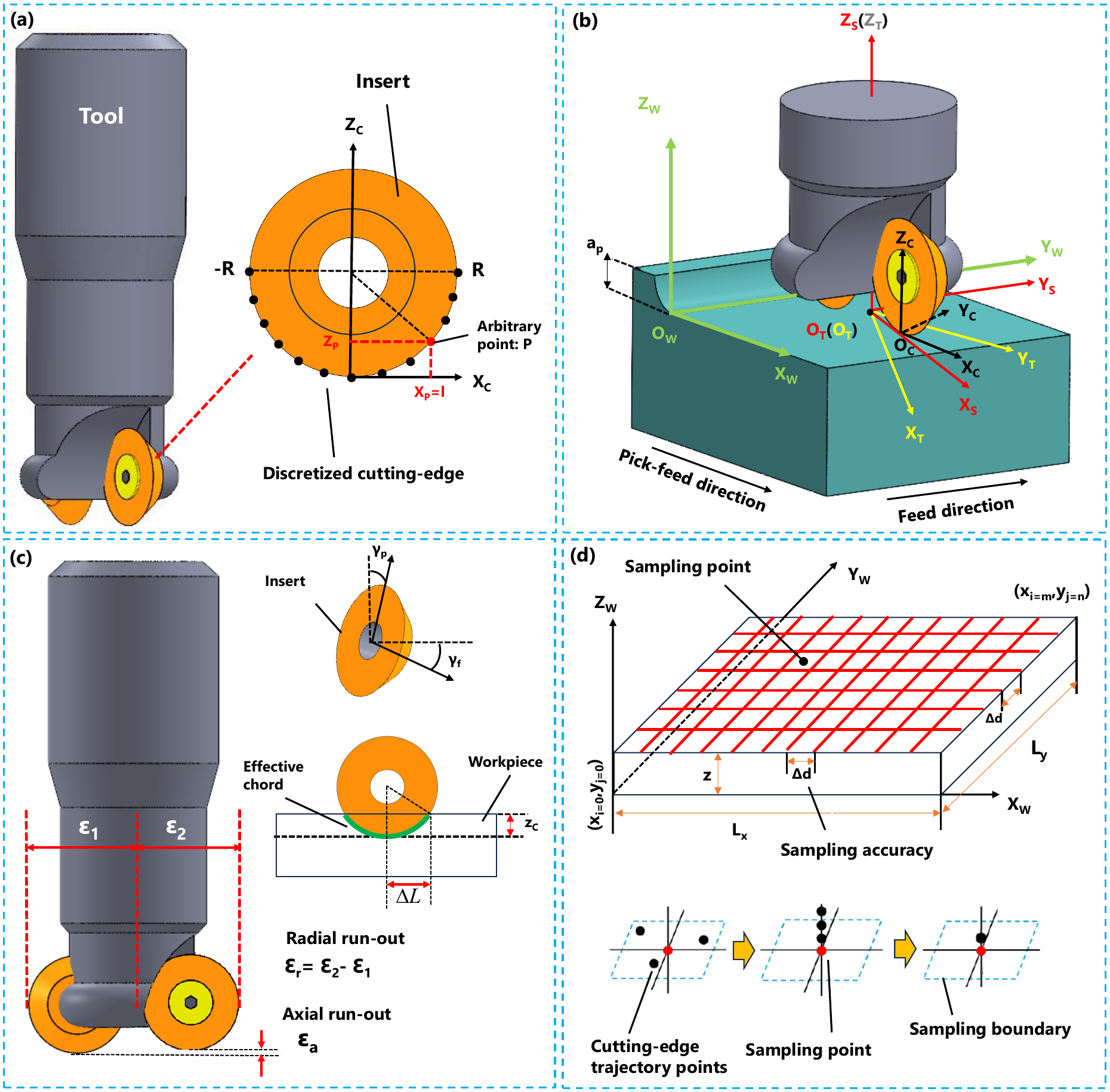}
            \caption{(a) Indexable face milling cutter with two inserts. The inserts have a radius $R$, and the interest point $P$ is located on the engaged cutting-edge. (b) Four coordinate systems are defined: the cutting-edge, tool, spindle, and workpiece coordinate systems. The point $P$ on the cutting-edge moves during the milling operation, and its trajectory is obtained in the workpiece coordinate system using the kinematic transformation equations between the coordinate systems. (c) Axial and radial run-outs, with the axial and radial rake angles of the tool. (d) Discretization of the workpiece. The workpiece is discretized into grids with sampling points located within each grid and along the sampling boundaries. The $z$ represents the depth of cut in the milling process. Additionally, the topography generation process within each sampling boundary is presented.}\label{Tool_Insert_Coordinates}
        \end{figure}

        To describe the motion of the cutting-edge, it is necessary to define the kinematic equations that relate the predefined coordinate systems. The main objective is to obtain the transformation equations between these coordinate systems for a point of interest $P$ on the cutting-edge, as illustrated in \autoref{Tool_Insert_Coordinates}a. In face milling with an indexable face mill cutter, four coordinate systems are typically defined \cite{wang2023high}, as shown in \autoref{Tool_Insert_Coordinates}b.
    
        The first is the cutting-edge coordinate system (${{O}_{C}}-{{X}_{C}}{{Y}_{C}}{{Z}_{C}}$), whose origin is located at the lowest point of the cutting-edge. The ${X}_{C}$ axis is aligned with the cutting-edge profile, ${Y}_{C}$ is normal to the cutting-edge, and ${Z}_{C}$ is oriented perpendicular to the workpiece surface. A point of interest on the cutting-edge is defined by the coordinates $({{x}_{P}},{{y}_{P}},{{z}_{P}})$. For circular inserts, the corresponding points along the active cutting-edge, the lower semicircular portion of the insert, can be expressed as
    
        \begin{equation}\label{eq_point_p_equation}
            \left\{ \begin{matrix}
           {{x}_{P}}=l\,,\,\,\,\,\,l\in [-R,\,R]  \\
           {{y}_{P}}=0\,\,\,\,\,\,\,\,\,\,\,\,\,\,\,\,\,\,\,\,\,\,\,\,\,\,\,\,\,\,\,\,  \\ {{z}_{P}}=R-\sqrt{{{R}^{2}}-{{l}^{2}}}\,\,\,  \\
            \end{matrix} \right.
        \end{equation}
    
        \noindent where $R$ denotes the insert radius. In this representation of a point on the cutting-edge, ${x}_{P}$ varies between the negative and positive values of the radius, ${y}_{P}$ is zero because it lies tangent to the cutting-edge, and ${z}_{P}$ varies as the point moves along the peripheral path of the lower semicircular portion of the cutting-edge. Due to the rotational and translational motions of the cutting-edge, these coordinates change and must be mapped to subsequent coordinate systems. 
        
        The second coordinate system is the tool coordinate system (${{O}_{T}}-{{X}_{T}}{{Y}_{T}}{{Z}_{T}}$), whose origin is located on the tool centerline at the same height as the lowest point of the cutting-edge. The ${X}_{T}$ and ${Y}_{T}$ axes rotate with the tool during operation, while the ${Z}_{T}$ axis remains normal to the workpiece surface. To transform the point $P$ from the cutting-edge coordinate system to the tool coordinate system, both the rotational and translational components must be considered. This is necessary because the axial and radial rake angles are involved, as well as the distance of the cutting-edge from the tool center, which depends on the tool diameter and the insert index. The transformation from the cutting-edge system to the tool coordinate system is denoted by ${{T}_{C\to T}}$ and can be expressed as
    
        \begin{equation}\label{eq_cutting_edge_to_tool}
            {{T}_{C\to T}}=\left[ \begin{matrix}
               \cos ({{\gamma }_{f}}) & \sin ({{\gamma }_{f}})\cos ({{\gamma }_{p}}) & \sin ({{\gamma }_{f}})\sin ({{\gamma }_{p}}) & \frac{D}{2}+(K-1){{\varepsilon }_{r}}  \\
               -\sin ({{\gamma }_{f}}) & \cos ({{\gamma }_{f}})\cos ({{\gamma }_{p}}) & \cos ({{\gamma }_{f}})\sin ({{\gamma }_{p}}) & 0  \\
               0 & -\sin ({{\gamma }_{p}}) & \cos ({{\gamma }_{p}}) & (K-1){{\varepsilon }_{a}}  \\
               0 & 0 & 0 & 1  \\
            \end{matrix} \right]
        \end{equation}
    
        In this transformation matrix, the first three columns describe the rotational behavior of the coordinate axes, where ${\gamma }_{f}$ represents the radial rake angle and ${\gamma }_{p}$ denotes the axial rake angle of the tool. The last column accounts for the translational components of the transformation. In this context, $D$ is the cutting diameter of the tool, $K$ is the index of the cutting-edge, ${\varepsilon }_{a}$ denotes the axial run-out of the inserts, and ${\varepsilon }_{r}$ represents the radial run-out of the inserts as shown in \autoref{Tool_Insert_Coordinates}c.
    
        The third coordinate system is the spindle coordinate system, denoted by (${{O}_{S}}-{{X}_{S}}{{Y}_{S}}{{Z}_{S}}$), which shares the same origin as the tool coordinate system. In this frame, ${X}_{S}$ defines the feed direction, ${Y}_{S}$ represents the step-over (pick-feed) direction, and the ${Z}_{S}$ axis is perpendicular to the workpiece surface. The transformation matrix from the tool coordinate system to the spindle coordinate system is denoted by ${{T}_{T\to S}}$ and can be expressed as
    
        \begin{equation}\label{eq_tool_to_spindle}
                {{T}_{T\to S}}=\left[ \begin{matrix}
               \cos \left( \varphi +\frac{2\pi (K-1)}{{{z}_{n}}}-\omega t \right) & \sin \left( \varphi +\frac{2\pi (K-1)}{{{z}_{n}}}-\omega t \right) & 0 & 0  \\
               -\sin \left( \varphi +\frac{2\pi (K-1)}{{{z}_{n}}}-\omega t \right) & \cos \left( \varphi +\frac{2\pi (K-1)}{{{z}_{n}}}-\omega t \right) & 0 & 0  \\
               0 & 0 & 1 & 0  \\
               0 & 0 & 0 & 1  \\
            \end{matrix} \right]
        \end{equation}
    
        \noindent where $\omega$ is the angular velocity of the tool, $\varphi$ is the initial phase angle, ${z}_{n}$ denotes the number of tool teeth (inserts), and $t$ represents the milling time. Transformation from the tool coordinate system to the spindle coordinate system involves only rotational components about the ${X}$ and ${Y}$ directions. These rotations incorporate the phase angle and angular velocity according to the relative position of the $K$th insert within the cutter assembly. The ${Z}$ axes remain unchanged because they are aligned, and no translational component is required since both coordinate systems share the same origin. 
    
        The final coordinate system is the workpiece coordinate system. The objective is to ultimately transform the coordinates of a point on the cutting-edge into the workpiece coordinate system in order to track the trajectory of the cutting-edge during machining. As illustrated in \autoref{Tool_Insert_Coordinates}b, the workpiece coordinate system (${{O}_{W}}-{{X}_{W}}{{Y}_{W}}{{Z}_{W}}$) is defined with its origin located on the machined surface of the workpiece. Its axes are aligned with those of the spindle coordinate system, ensuring consistency between the two reference frames. This alignment implies that only translational motion exists between these coordinate systems, which can be expressed as
    
        \begin{equation}\label{eq_spindle_to_workpiece}
                {{T}_{S\to W}}=\left[ \begin{matrix}
               1 & 0 & 0 & {{x}_{0}}  \\
               0 & 1 & 0 & {{y}_{0}}+{{v}_{f}}t  \\
               0 & 0 & 1 & {{z}_{0}}  \\
               0 & 0 & 0 & 1  \\
                \end{matrix} \right]
        \end{equation}
    
        \noindent where ${v}_{f}$ denotes the feed speed, and $({{x}_{0}},{{y}_{0}},{{z}_{0}})$ represents the initial position of the tool in the workpiece coordinate system. As can be observed, the translational component of the matrix describes the motion of the tool in the feed direction.
        
        Ultimately, by multiplying the transformation matrices, the coordinates of the point $P$ in the workpiece coordinate system can be obtained, i.e., $({{{x}'}_{P}},{{{y}'}_{P}},{{{z}'}_{P}})$, according to the following equation.
        
        \begin{equation}\label{eq_full_transformation}
                \left[ \begin{matrix}
               {{{{x}'}}_{P}}  \\
               {{{{y}'}}_{P}}  \\
               {{{{z}'}}_{P}}  \\
               1  \\
            \end{matrix} \right]={{T}_{S\to W}}{{T}_{T\to S}}{{T}_{C\to T}}\left[ \begin{matrix}
               {{x}_{P}}  \\
               {{y}_{P}}  \\
               {{z}_{P}}  \\
               1  \\
            \end{matrix} \right]
        \end{equation}
    
        \autoref{eq_full_transformation} provides the coordinates of each point on the cutting-edge of a given tooth under specific milling conditions in the workpiece coordinate system. Therefore, it defines the path of the cutting-edge throughout the operation.

    \subsection{Discretization process}
        \label{subsec_discretization_process}

        By deriving the transformation equation of an arbitrary point located on the cutting-edge, its trajectory within the workpiece can be determined through spatial discretization of the edge into a finite number of points. The trajectory of each discretized point is subsequently evaluated in the workpiece coordinate system.
        
        Within the framework of the FSM, three principal components require discretization. The first component is the cutting-edge. As illustrated in \autoref{Tool_Insert_Coordinates}a, the cutting-edge is discretized into a series of points where their coordinates are defined according to \autoref{eq_point_p_equation}.
        
        In order to optimize the model, this discretization can be refined by computing the actual chord length of the cutting edge that interacts with the workpiece. This discretization is performed along the local edge coordinate $X_{C}$, which lies in the radial-axial plane. The discretization length $\Delta L$ is defined based on two criteria: the axial depth of cut $a_p$ and the feed per tooth $f_z$.
        
        First, the half-chord length associated with the axial immersion $z_{C}$ is given by $\Delta L_{a_p} = \sqrt{R^2 - (R - z_{C})^2}$ as shown in \autoref{Tool_Insert_Coordinates}c.
        
        Second, a minimum discretization length is imposed based on the feed per tooth, defined as $\Delta L_{f_z} = \frac{f_z}{2 \cos\gamma_f}$, where $\gamma_f$ the rake angle, and $f_z$ the feed per tooth. 
        
        The effective half-length used for discretization is then defined as $\Delta L = \max\left( \Delta L_{a_p},\, \Delta L_{f_z} \right)$. If the cutting edge is discretized into $N$ points uniformly distributed along the interval $[-\Delta L,\, \Delta L]$, then the coordinate of each point is given by $l_{P} = -\Delta L + \frac{2 \Delta L}{N-1} , {P}$, with ${P} = 0, \dots, N-1$.
        
        For each discretized point, the corresponding axial coordinate is obtained from the circular profile of the tool as $z_{P} = R - \sqrt{R^2 - l_{P}^2}$. Thus, each point of the cutting edge in homogeneous coordinates is expressed as

        \begin{equation}\label{eq_tool_point_p_equation}
            \left\{ \begin{matrix}
           {{x}_{P}}=l_{P}\,,\,\,\,\,l_{P}\in [-\Delta L,\,\Delta L]  \\
           {{y}_{P}}=0\,\,\,\,\,\,\,\,\,\,\,\,\,\,\,\,\,\,\,\,\,\,\,\,\,\,\,\,\,\,\,\,\,\,\,\,\,\,\,\,\,\,\,\,\,\,  \\ 
           {{z}_{P}}=R-\sqrt{{{R}^{2}}-{{l_{P}}^{2}}}\,\,\,\,\,\,\,\,\,\,\,\,\,  \\
            \end{matrix} \right.
        \end{equation}
        
        This discretization ensures that the cutting edge representation captures both the tool geometry and the kinematic constraints imposed by the feed motion.
        
        The second component is the workpiece surface, which is discretized as shown in \autoref{Tool_Insert_Coordinates}d. The surface is partitioned using a sampling interval of $\Delta d$ along the ${X}_{W}$ and ${Y}_{W}$ directions. No discretization is applied along the ${Z}_{W}$ direction; instead, this coordinate corresponds to the prescribed depth of cut. Consequently, the discretized surface consists of $(m+1)\times(n+1)$ grid nodes indexed by $(i,j)$, each enclosed by defined sampling boundaries.
        
        Finally, temporal discretization is introduced, as the kinematic behavior of all discretized points is time-dependent. A sequence of discrete time increments, denoted by $\Delta t$, is considered, and the governing equations are evaluated at each time step to accurately capture the evolution of the cutting process.

    \subsection{3D surface topography generation}
        \label{subsec_3D_surface_topography}

        During the simultaneous translational and rotational motion of the milling tool in the face-milling operation, discretized points along the cutting-edge generate spatial motion trajectories. By substituting the discretization intervals of the cutting-edge and time in \autoref{eq_full_transformation}, together with the relevant milling parameters, the complete kinematic description of each point is obtained. Therefore, a large number of trajectory points are generated throughout the process. As these points intersect the discretized workpiece surface, their positions at each time step fall within specific sampling boundaries of the workpiece grid. This positional relationship can be expressed as

        \begin{equation}\label{eq_sampling_boundaries}
            \left\{ \begin{matrix}
           {{x}_{i}}-\frac{\Delta d}{2}\le {{{{x}'}}_{p}}\le {{x}_{i}}+\frac{\Delta d}{2}  \\
           {{y}_{j}}-\frac{\Delta d}{2}\le {{{{y}'}}_{p}}\le {{y}_{j}}+\frac{\Delta d}{2}  \\{{{{z}'}}_{p}}\,\,\,\,\,\,\,\,\,\,\,\,\,\,\,\,\,\,\,\,\,\,\,\,\,\,\,\,\,\,\,\,\,\,\,\,\,\,\,\,\,\,\,\,\,\,\,\,  \\
            \end{matrix} \right.
        \end{equation}

        \noindent where $({{{x}'}_{p}},{{{y}'}_{p}},{{{z}'}_{p}})$ denotes the trajectory point of the cutting-edge expressed in the workpiece coordinate system. As illustrated in \autoref{Tool_Insert_Coordinates}d, the cutting-edge trajectory points are distributed within the sampling boundaries of the discretized workpiece surface grid. For each specific sampling boundary, the minimum value of the height coordinate $z$ among all trajectory points falling within that boundary is selected to represent the local surface topography. Through this procedure, retaining the lowest $z$ value within each sampling boundary progressively generates the machined surface topography as a cloud of points over the workpiece surface, obtained by traversing all computed trajectory points.
        
\begin{figure}[ht]%
            \centering
            \includegraphics[width=0.9\textwidth]{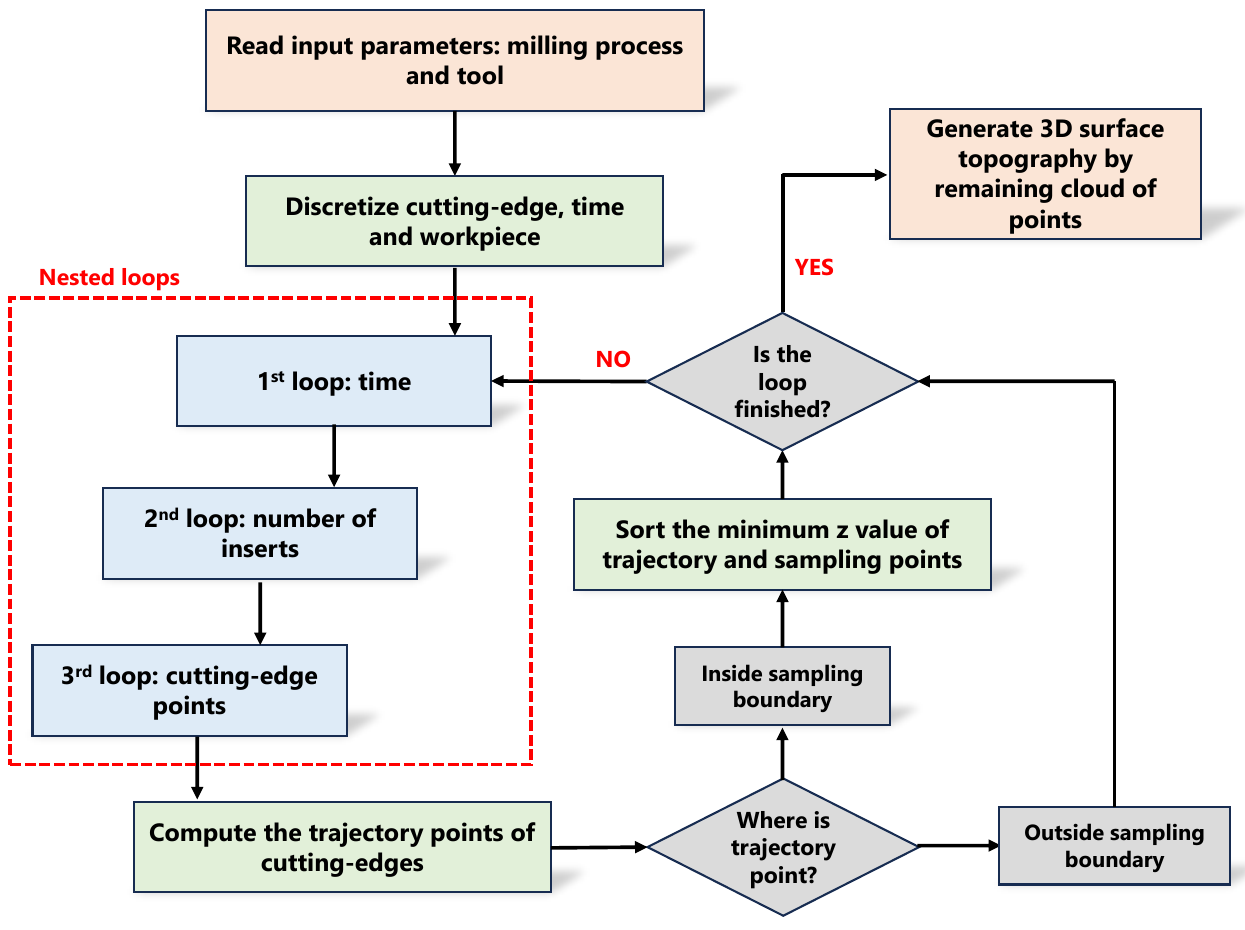}
            \caption{Flowchart of the FSM algorithm.}\label{FSM_Flowchart}
        \end{figure}

        The overall flowchart of the FSM is depicted in \autoref{FSM_Flowchart}. The algorithm begins with the specification of the milling and tool parameters. Subsequently, the discretization process is performed, including spatial discretization of the cutting-edge and the workpiece surface, as well as temporal discretization of the time domain. The core of the algorithm consists of nested loops over all discretized spatial and temporal elements. Within these loops, the trajectory points are computed, and whenever a point lies within a given sampling boundary, its $z$ value is compared with the existing value assigned to that boundary. The minimum value is retained as the representative surface point. After completion of all iterations, the resulting cloud of retained points is used to reconstruct and visualize the 3D surface topography.

        The FSM algorithm shown in \autoref{FSM_Flowchart} relies heavily on deeply nested loops, which significantly degrade computational performance due to repeated iterations over numerous spatial and temporal steps. This structure generates a substantial number of repetitive operations. In Python, as an interpreted language, each iteration of a $for$ loop is executed at the interpreter level, introducing overhead at every step. Consequently, millions of small operations accumulate considerable execution time. 
        
        This limitation becomes particularly pronounced when very small time steps, large surface areas, or high spatial resolutions are required, as both conditions drastically increase the total number of iterations. As a result, the algorithm scales poorly and incurs significant computational cost. These performance constraints necessitate the exploration of more efficient computational frameworks and optimization strategies to enhance scalability and overall efficiency. The developed interpreter-based implementation of the FSM algorithm is publicly available on GitHub at \url{https://github.com/HadiBakhshan/surf-topo.git}.

\section{Development of the efficient framework}
    \label{sec_development_framework}

    Computational efficiency is generally influenced by both data volume and the underlying model algorithm \cite{wang2023high}. However, in milling process simulations, scenarios such as large surface areas, low feed rates, or very fine discretization of the tool and workpiece inevitably lead to substantial data generation. In such cases, control over data volume becomes limited, and performance improvements must primarily be achieved through algorithmic optimization.
    
    Therefore, this section introduces an optimized and computationally efficient framework based on the original FSM, referred to as the efficient FSM (EFSM). The proposed EFSM aims to improve the computational performance of the FSM by reducing computational cost and improving overall efficiency while maintaining modeling accuracy.

    \subsection{General architecture of EFSM}
        \label{subsec_general_architecture_EFSM}

        The EFSM framework is developed using a hybrid C++/Python architecture to combine computational efficiency with high-level usability. In this structure, the core numerical engine is implemented in C++, while Python is employed for configuration, execution control, and post-processing tasks. In milling process simulations for surface topography, where deeply nested loops over spatial and temporal variables are required, this separation becomes essential to avoid the significant overhead associated with interpreted languages.
        
        As a compiled language, C++ translates source code into optimized machine instructions prior to execution. This enables tight loops, matrix operations, and memory indexing routines to execute with minimal runtime overhead. Moreover, C++ provides explicit control over memory allocation, contiguous data storage, and object lifetimes, thereby improving cache locality and reducing dynamic allocation costs during simulation. These advantages become particularly critical in high-resolution simulations involving a large number of discretized points.

        \begin{figure}[ht]%
            \centering
            \includegraphics[width=0.99\textwidth]{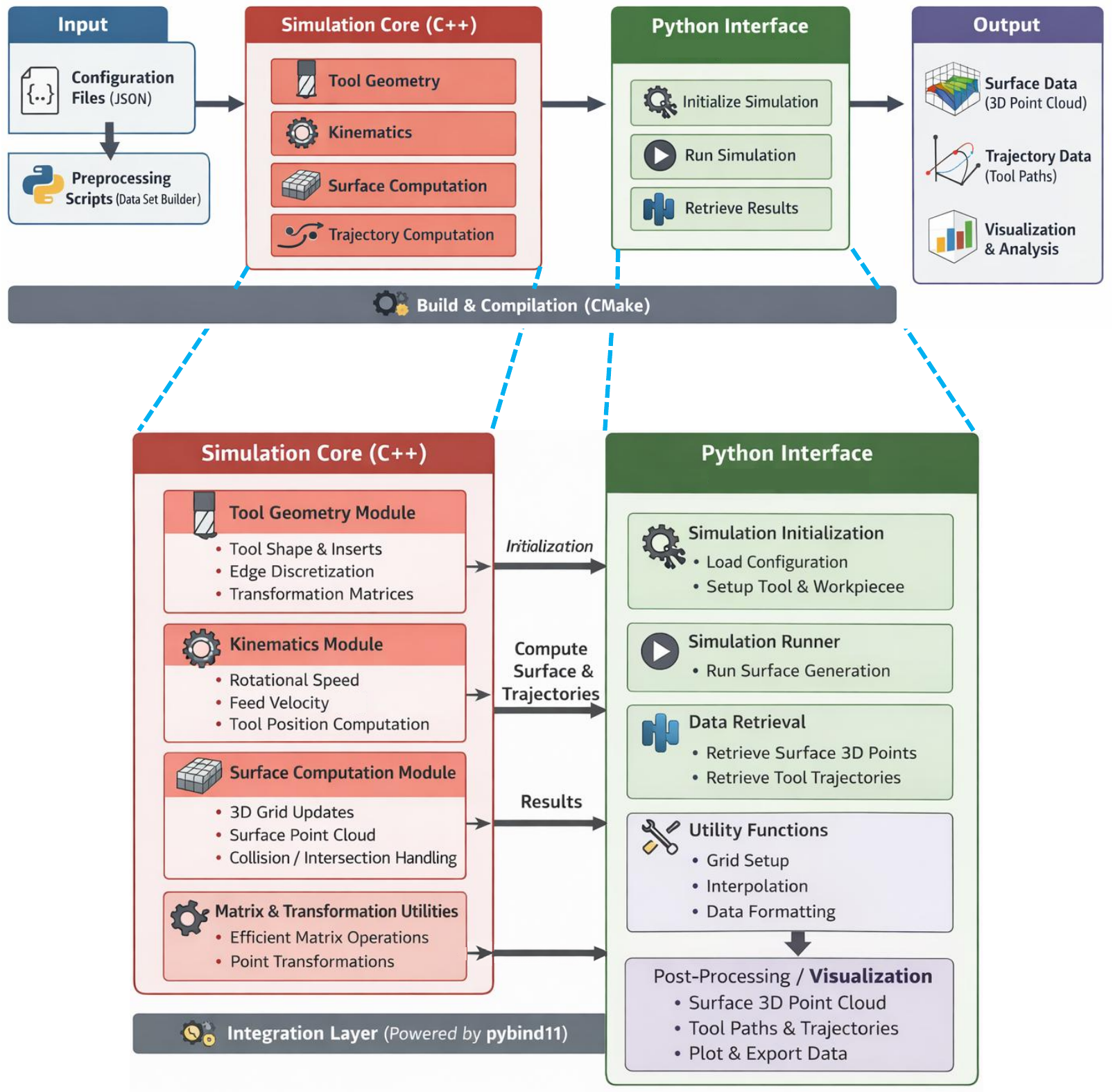}
            \caption{Architecture of the EFSM.}\label{General_Architecture_EFSM}
        \end{figure}

        Within the EFSM framework, all computationally intensive tasks such as geometric transformations, surface height updates, and trajectory calculations are executed entirely within the C++ domain. \autoref{General_Architecture_EFSM} illustrates the general architecture of the EFSM, which consists of four main components: input, simulation core, Python interface, and output. The framework follows a layered architecture in which the computational core is implemented in C++, while orchestration, configuration, and post-processing are handled through a Python interface connected via \texttt{pybind11} \cite{pybind11}. The binding layer exposes the C++ classes and methods as Python-callable objects. Constructors, simulation execution functions, and data accessors are registered explicitly, preserving type safety at compile time. Numerical outputs are transferred to Python using \texttt{NumPy}-compatible \cite{harris2020array} buffer protocols, which minimizes unnecessary copying of large arrays.

\begin{figure}[ht]%
            \centering
            \includegraphics[width=0.6\textwidth]{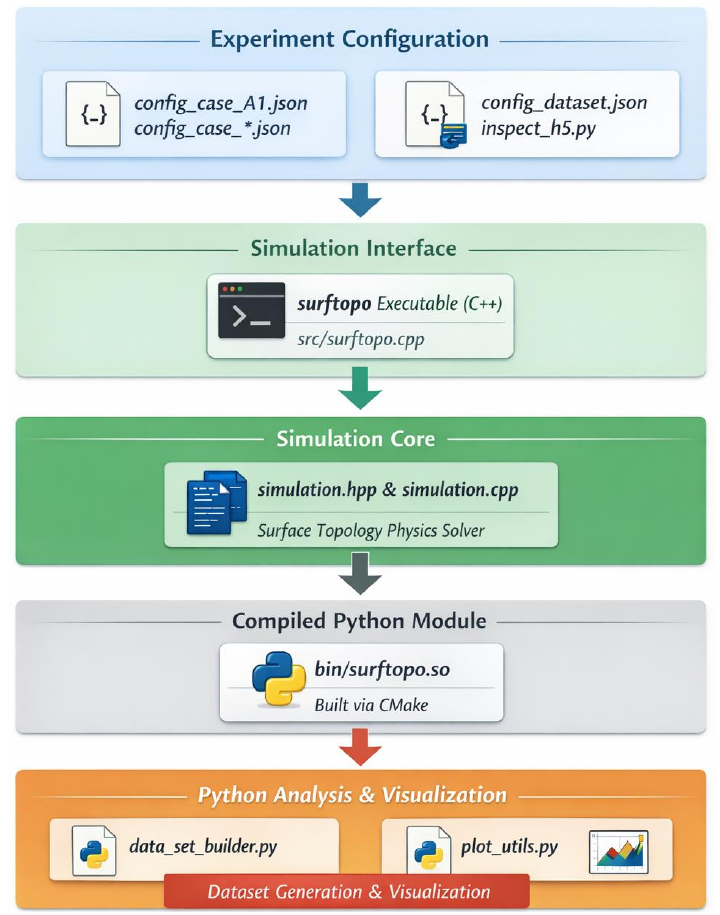}
            \caption{Overview of the \textit{surf-topo} framework workflow. Simulation cases are specified using JSON configuration files that define individual experiments and dataset settings. These configurations are executed by the C++ simulation interface (\texttt{surftopo.cpp}), which invokes the core solver implemented in \texttt{simulation.hpp} and \texttt{simulation.cpp}. The solver is compiled via \texttt{CMake} into a Python-accessible module ( \texttt{surftopo.so}). Python utilities are then used for dataset generation, inspection, and visualization of simulation results.}\label{Repository_Achitecture}
        \end{figure}

        The input part incorporates configuration files in JSON format containing all required parameters, including tool geometry and milling process parameters provided by the user. These JSON files are parsed by Python scripts, which then interface with the computational core. The output part leverages Python visualization libraries such as \texttt{Matplotlib} \cite{Hunter:2007} for surface topography plotting and analysis. The source code building and compilation process is conducted using \texttt{CMake}. \autoref{Repository_Achitecture} illustrates the workflow of the framework, referred to as \textit{surf-topo}. The implementation of the proposed efficient framework is publicly available on GitHub at \url{https://github.com/HadiBakhshan/surf-topo.git}, along with comprehensive documentation and detailed compilation guidelines.

    \subsection{Simulation core}
        \label{subsec_simulation_core}

        The simulation component of the EFSM framework consists of four structural modules, as illustrated in \autoref{General_Architecture_EFSM}: tool geometry, kinematics, surface computation, and utilities.

        The tool geometry module is responsible to construct the discrete representation of the cutting tool prior to time integration. From a technical perspective, it initializes the geometric parameters such as tool diameter, insert radius, number of teeth, and rake angles, and performs edge discretization by sampling the insert profile into a finite set of points stored in contiguous memory containers (e.g., \texttt{std::vector}). These points are expressed in homogeneous coordinates to facilitate efficient transformation chaining during kinematic computations. In addition, this module precomputes constant geometric quantities and required trigonometric terms that are repeatedly used throughout the simulation. By evaluating these quantities once during initialization, redundant computations inside the iterative loops are avoided, thereby improving runtime efficiency. Furthermore, for geometric datasets such as the cutting-edge point cloud, memory is allocated during object construction and reused throughout execution. This design promotes cache-friendly memory access patterns and eliminates repeated heap allocations during runtime, contributing to improved computational performance and stability.

        The kinematics module governs the time-dependent motion of each tooth and is executed within the main simulation loop. For every discrete time step, it computes the instantaneous angular position of each cutting-edge and combines it with the translational feed motion according to \autoref{eq_full_transformation}. This module is therefore responsible for mapping point coordinates from the local cutting-edge reference frame to the global workpiece coordinate system. To implement these transformations, fixed-size $4\times 4$ homogeneous transformation matrices are employed. Operations are carried out using stack-allocated arrays or lightweight matrix utilities to eliminate dynamic memory allocation overhead. This design ensures efficient execution of the repeated transformation operations within the time-stepping loop while maintaining numerical robustness.

        The surface computation module constitutes the dominant computational workload of the EFSM framework. It maintains the structured workpiece surface grid represented as a height field, typically stored in a flattened contiguous array to enhance spatial locality and cache efficiency. For each transformed tool point provided by the kinematics module, the corresponding grid indices are computed using integer arithmetic derived from the predefined discretization resolution. Regarding the material removal mechanism, a conditional minimum operation is applied: if the $z$-coordinate of the transformed tool point is lower than the currently stored grid height, the grid value is overwritten. This update procedure is executed within tightly nested loops over time steps, teeth, and discretized edge points. However, since the implementation is fully compiled in C++, it avoids the interpreter overhead that would be incurred in an equivalent Python-based approach. To ensure robustness, explicit boundary checks and index clamping are performed to prevent out-of-range memory access. Moreover, to guarantee deterministic runtime behavior and stable memory usage, no dynamic memory allocation occurs within these computational loops; all required arrays are preallocated prior to entering the simulation phase.
        
        The trajectory computation module, as depicted in the detailed flowchart, operates in parallel with the surface updating procedure while maintaining a logically separate data structure. At each time step, this module identifies the minimum $z$-value along each discretized cutting-edge and records its corresponding global coordinates. The storage container for trajectory data is typically reserved in advance to prevent repeated resizing during execution. This separation of trajectory tracking from surface updating preserves modular clarity, while both modules share the same transformation data generated by the kinematics module.

        The matrix and transformation utilities module supports the computational modules described above, particularly in implementing the transformation operations. This utility layer provides reusable routines for homogeneous matrix construction, matrix multiplication, and coordinate mapping between reference frames. By centralizing transformation-related operations, the framework avoids code duplication and promotes maintainability while enabling potential low-level optimizations, such as function inlining and vectorized arithmetic. The module is purely computational in nature and does not manage persistent state beyond temporary transformation objects required during execution. This design ensures modularity, clarity, and computational efficiency within the overall EFSM architecture.

    \subsection{Python interface}
        \label{subsec_python_interface}

        The Python Interface constitutes another key component of the framework and is responsible for high-level orchestration of the simulation workflow. It initializes simulation parameters and instantiates the underlying C++ simulation objects.
        
        The simulation initialization block corresponds to Python-side object construction and parameter configuration. Configuration parameters are read (e.g., from JSON files), validated, and then passed to the C++ simulation classes through \texttt{pybind11} bindings. At this stage, all parameters are transferred to the C++ domain, where geometric entities and surface grid structures are allocated and initialized. Importantly, after this setup phase, no computationally intensive loops remain in Python.
        
        The simulation runner acts as a lightweight wrapper around the C++ execution method. Once invoked, the control is fully transferred to the C++ simulation core. All iterative operations including time stepping, tooth indexing, geometric transformations, and surface updates are executed entirely within the compiled environment. This strict separation ensures that no cross-language function calls occur within performance-critical loops, thereby eliminating binding overhead during execution.
        
        Following simulation completion, the data retrieval stage is handled through accessor functions exposed via \texttt{pybind11}. These functions transfer the computed surface and trajectory arrays from C++ memory into \texttt{NumPy}-compatible data structures. Furthermore, data ownership semantics are explicitly defined to ensure that memory remains valid while referenced by Python objects, thereby maintaining both safety and performance.

% \section{Case study}
%     \label{sec_case_study}

\section{Model validation}
    \label{sec_model_validation}

    This section evaluates the validity of the proposed model by assessing its accuracy through comparison with two independent sets of experimental results. The first set comprises newly conducted milling experiments performed under two distinct operating conditions, while the second set reproduces and compares results reported in the literature. The computational performance of the model is further assessed using this second set of literature-based experiments. Through these comparisons, the general applicability and robustness of the EFSM across different tools and machining conditions can be effectively evaluated.

    \subsection{Case 1: newly conducted milling experiments}
        \label{subsec_case_1}

        \subsubsection{Experimental procedure}
            \label{subsubsec_case_1_experimental_procedure}

            The workpiece selected for the milling process experiments was fabricated from PH17-4 stainless steel using laser wire directed energy deposition (DED), as this material is readily available in the required wire format for the DED process. A layer height of 1 mm was employed, and a raster deposition pattern was used to build the workpiece with a wire feed speed of 15 mm/s.

\begin{figure}[ht]%
                \centering
                \includegraphics[width=0.99\textwidth]{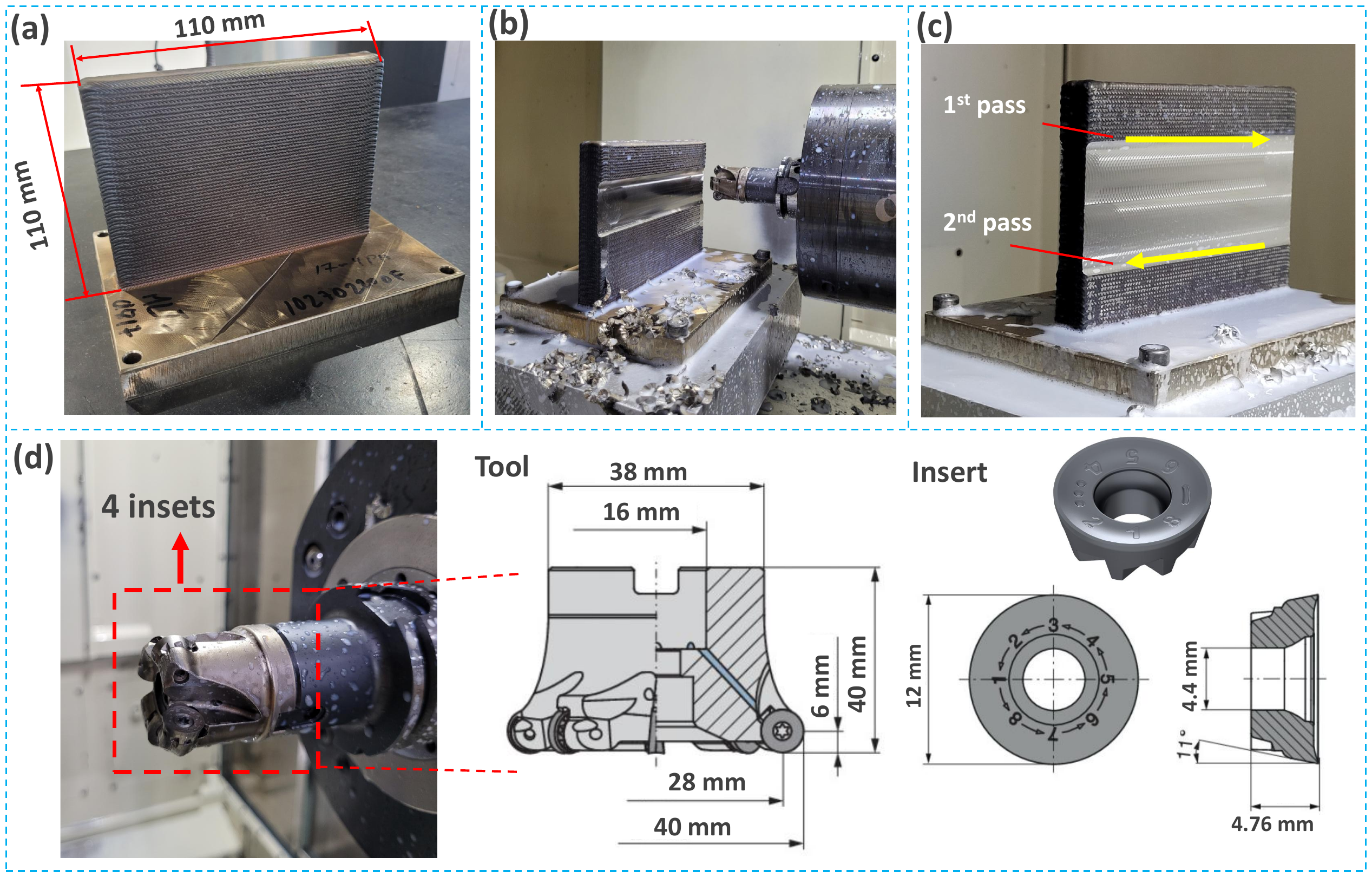}
                \caption{Experimental setup. (a) Additively manufactured workpiece and its dimensions. (b) Face milling process. (c) Two passes of the tool on the workpiece during the milling operation. (d) Face mill cutter used in the operation, equipped with four inserts, along with the tool and insert specifications.}\label{Expeimental_Setup}
            \end{figure}

            Heat treatment was performed on the fabricated workpieces through solution annealing at ${{1040}^{\circ }}$C for 30 minutes, followed by aging at ${{550}^{\circ }}$C for 4 hours, resulting in a final hardness of approximately 35 HRC. The dimensions of the additively manufactured workpiece were $110\times 110$ mm, as shown in \autoref{Expeimental_Setup}a.

            Face milling operations were carried out using a four-insert cutter, as illustrated in \autoref{Expeimental_Setup}b, under two machining conditions, namely smooth and aggressive. The detailed milling parameters are summarized in \autoref{tab_milling_paramters_additively}. The milling operations were performed in two passes, as shown in \autoref{Expeimental_Setup}c. A MaxiMill 251 RS face mill equipped with RPX 1204 inserts was used for all milling trials, with further tool specifications provided in \autoref{Expeimental_Setup}d. Topographic images were acquired using a non-contact 3D profilometer, the ALICONA Infinite Focus G5, with an S-filter of 10 $\mu$m and an L-filter of 0.8 mm. This optical profilometer enables accurate roughness measurements and high-resolution characterization of surface morphology.

            \begin{table}[h]
                \centering
                \fontsize{8}{13}\selectfont
                \caption{Milling process parameters of additively manufactured parts.}
                \label{tab_milling_paramters_additively}
                \begin{tabularx}{0.65\textwidth}{>{\centering\arraybackslash}p{3cm} 
                                        >{\centering\arraybackslash}p{3cm} 
                                        >{\centering\arraybackslash}X 
                                        >{\centering\arraybackslash}X}
                    \toprule
                    Process & Spindle speed (rpm) & ${v}_{f}$ (mm/min) & ${{a}_{P}}$ (mm) \\ \toprule
        
                    Smooth & 995 & 125 & 2.5 \\
        
                    Aggressive & 995 & 1990 & 0.5 \\
           
                    \bottomrule
                \end{tabularx}
            \end{table}

        \subsubsection{Predictive performance}
            \label{subsubsec_case_1_Predictive}

            The predictability of the EFSM is primarily characterized using 3D surface topography, 2D roughness maps, and surface roughness metrics. The predicted results obtained from FSM and EFSM are essentially equivalent, since both methods employ the same discretization and computational procedures. Therefore, they exhibit comparable accuracy in terms of prediction performance.

\begin{figure}[ht]%
                \centering
                \includegraphics[width=0.98\textwidth]{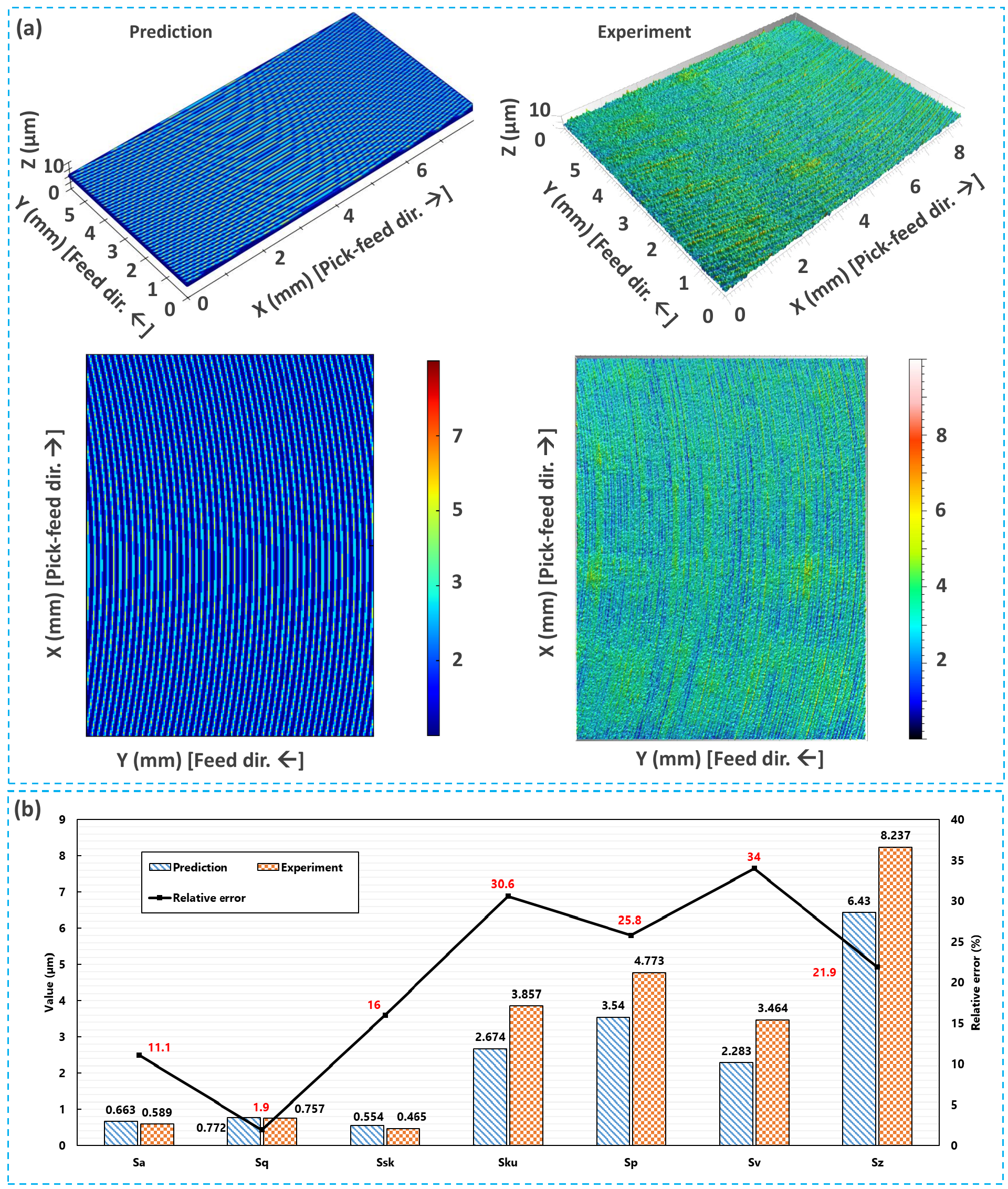}
                \caption{(a) 3D surface topographies and 2D roughness maps of the experimental measurements and corresponding predictions under smooth milling conditions. (b) Comparison of the predicted roughness metrics with their experimental counterparts in smooth milling conditions.}\label{3D_Surface_Topography_Smooth}
            \end{figure}

            For the milling process under smooth conditions, \autoref{3D_Surface_Topography_Smooth}a presents the 3D surface topography of the machined specimens together with the corresponding predicted surfaces, along with 2D roughness maps of the images. The contour distribution values in both the 3D and 2D representations demonstrate a close agreement between measurement and prediction. Moreover, the surface patterns generated on the workpiece, resulting from the tool trajectories and feed motion, show similar characteristics. It can be observed that, due to the low feed rate, ordered and symmetric curved tool marks remain visible, indicating a regular sequence of peaks and valleys on the machined surface.

            In addition to the 3D and 2D representations of the roughness distribution on the workpieces, commonly used quantification parameters are employed to better evaluate the predictive performance of the proposed EFSM model. In this regard, the standardized areal surface texture parameters defined in ISO 25178-2 \cite{ISO25178-2-2021} are considered. Among these, the most widely reported metric is the arithmetic mean height or average surface roughness (${S}_{a}$), which represents the average absolute deviation of surface heights from the mean plane. The root mean square height (${S}_{q}$) is also frequently used, as it is more sensitive to high peaks and deep valleys due to the squaring operation. These metrics are categorized as amplitude parameters. However, functional and spatial parameters such as the maximum peak height (${S}_{p}$), maximum valley depth (${S}_{v}$), maximum height of the surface (${S}_{z}$, defined as the distance between the lowest valley and the highest peak), skewness (${S}_{sk}$), and kurtosis (${S}_{ku}$) provide a more comprehensive characterization of the surface topography compared to traditional profile-based roughness parameters.

            \autoref{3D_Surface_Topography_Smooth}b illustrates the measured and predicted values of these roughness metrics. Observation of the parameters indicates that, for average roughness prediction (such as ${S}_{a}$ and ${S}_{q}$), the relative errors remain low. However, for the prediction of extreme height parameters (i.e., maximum peak height, maximum valley depth, and their difference ${S}_{z}$), the prediction error increases slightly. The relative error in predicting the highest peak and deepest valley is greater compared to that of the average roughness metrics. This discrepancy may arise from the epistemic limitations of the computational framework, which is primarily based on geometrical tool-workpiece interaction modeling, whereas the actual machined surface is influenced by additional factors such as material heterogeneity, tool wear, vibration, and process dynamics. Nevertheless, the overall predictive performance can be considered acceptable on average, particularly for the key roughness indicators used in practical surface quality assessment.

\begin{figure}[ht]%
                \centering
                \includegraphics[width=0.98\textwidth]{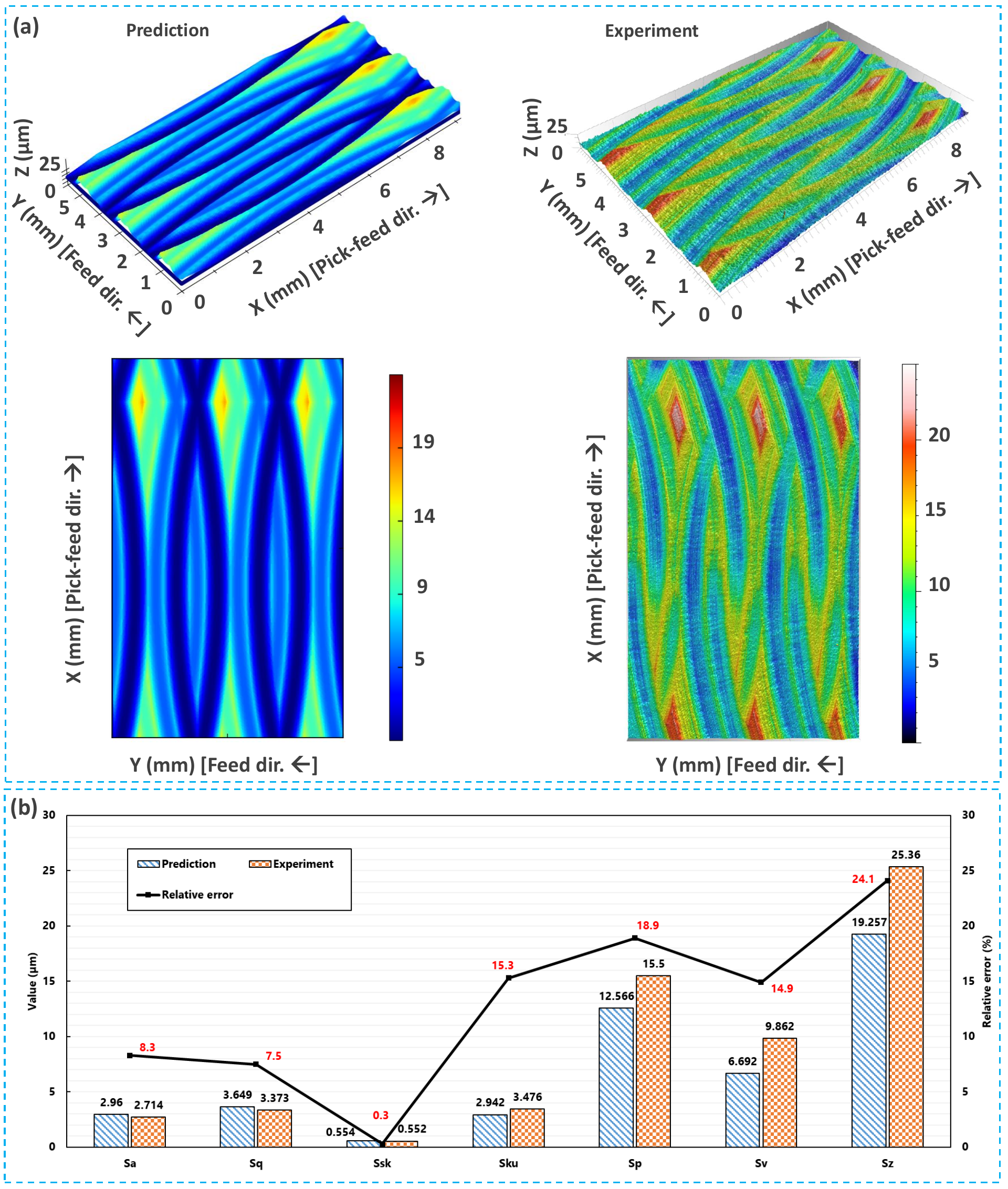}
                \caption{(a) 3D surface topographies and 2D roughness maps of the experimental measurements and corresponding predictions under aggressive milling conditions. (b) Comparison of the predicted roughness metrics with their experimental counterparts in aggressive milling conditions.}\label{3D_Surface_Topography_Aggressive}
            \end{figure}

            \autoref{3D_Surface_Topography_Aggressive}a shows the 3D surface topography and 2D roughness maps for the aggressive milling condition. It is evident that the surface patterns differ significantly from those obtained under smooth milling conditions. In the aggressive case, the feed rate is considerably higher, resulting in distinct tool-mark patterns and larger roughness contour values. This indicates that increasing the feed rate causes faster tool movement across the workpiece, thereby reducing the tool-surface interaction time per unit area, which ultimately leads to a harsher and rougher surface finish.

            This observation is further supported by the roughness metrics presented in \autoref{3D_Surface_Topography_Aggressive}b, where higher values of the average roughness parameters ${S}_{a}$ and ${S}_{q}$ can be observed. The predicted values show good agreement with the experimental measurements, with relatively low prediction errors. In addition, the prediction errors associated with the maximum and minimum height parameters are slightly higher than those observed under smooth milling conditions.

            Furthermore, surface skewness (${S}_{sk}$) , which represents the asymmetry of the surface height distribution relative to the mean plane, shows positive values in both the experimental and predicted results. This suggests a surface dominated by pronounced peaks, often referred to as a spiky texture. In contrast, negative skewness values would indicate a surface characterized by deeper valleys or plateau-like features.

            On the other hand, surface kurtosis (${S}_{ku}$) describes the sharpness or peakedness of the surface height distribution. A value of ${S}_{ku}=3$ corresponds to a normal (Gaussian) distribution. Values greater than 3 indicate a surface with sharp peaks and deep valleys (leptokurtic distribution), reflecting the presence of extreme surface features, whereas values lower than 3 represent a flatter height distribution (platykurtic surface).

            In this study, the experimental values of ${S}_{ku}$ are greater than 3, while the predicted values tend to be lower than 3, indicating that the model predicts a more moderate height distribution compared to the actual machined surface. This suggests that although the overall predictive capability of the model is robust, particularly for average roughness indicators, its performance is comparatively lower when predicting the detailed characteristics of extreme peaks and valleys. Such discrepancies are expected, as accurate prediction of localized surface features is inherently more challenging than predicting global roughness trends.

    \subsection{Case 2: literature-based experimental results}
        \label{subsec_case_2}

        \subsubsection{Experimental procedure}
            \label{subsubsec_case_2_experimental_procedure}

            To further evaluate the robustness and generalization capability of the proposed model, additional experimental results from milling operations are examined in this section. In particular, face milling cases reported in \cite{wang2023high}, conducted under different cutting parameter combinations, were adopted to provide an independent validation of the model performance. The milling experiments were performed on HT300 grey cast iron workpieces with dimensions of $20\times 20\times 100\,m{{m}^{3}}$ using a face mill cutter. The corresponding tool and milling parameters are summarized in \autoref{tab_tool_paramters} and \autoref{tab_milling_paramters}, respectively. Prior to the experimental procedures, the run-outs of the inserts (illustrated in \autoref{Tool_Insert_Coordinates}c) were measured and reported in \autoref{tab_run_out_inserts}. Milling under dry conditions was carried out on an EM800A vertical machining center, and the WYKO NT9300 optical profiler was employed to capture the 3D surface topography of the machined surfaces \cite{wang2023high}.

    \begin{table}[h]
        \centering
        \fontsize{8}{13}\selectfont
        \caption{Tool parameters \cite{wang2023high}.}
        \label{tab_tool_paramters}
        \begin{tabularx}{0.5\textwidth}{X X}
            \toprule
            Parameter & Value \\ \toprule

            Tool & R217.29-2520.3-05.2.070 \\

            Cutting diameter & 10.0 mm \\

            Cutting diameter maximum & 20.0 mm \\

            Shank length & 66.0 mm \\

            Shank diameter & 25.0 mm \\

            Radial rake angle & ${{0.6}^{\circ }}$ \\

            Axial rake angle & ${{0.0}^{\circ }}$ \\

            Tooth number & 2 \\

            Insert & RDHW10T3M0-8-MD04 F40M \\

            Insert thickness & 3.97 mm \\

            Insert diameter & 10.0 mm \\

            Corner radius & 5.0 mm \\

            Coating technology & PVD \\
   
            \bottomrule
        \end{tabularx}
    \end{table}

    \begin{table}[h]
        \centering
        \fontsize{8}{13}\selectfont
        \caption{Milling process parameters \cite{wang2023high}.}
        \label{tab_milling_paramters}
        % \begin{tabularx}{0.5\textwidth}{X X X X}
        \begin{tabularx}{0.5\textwidth}{>{\centering\arraybackslash}X 
                                >{\centering\arraybackslash}X 
                                >{\centering\arraybackslash}X 
                                >{\centering\arraybackslash}X}
            \toprule
            No. & ${{v}_{c}}$ (m/min) & ${{f}_{z}}$ (mm/tooth) & ${{a}_{P}}$ (mm) \\ \toprule

            Case 1 & 170 & 0.6 & 0.5 \\

            Case 2 & 200 & 0.4 & 0.4 \\

            Case 3 & 230 & 0.5 & 0.3 \\
   
            \bottomrule
        \end{tabularx}
    \end{table}

    \begin{table}[h]
        \centering
        \fontsize{8}{13}\selectfont
        \caption{Run-out of inserts \cite{wang2023high}.}
        \label{tab_run_out_inserts}
        % \begin{tabularx}{0.4\textwidth}{X X X}
        \begin{tabularx}{0.4\textwidth}{>{\centering\arraybackslash}X 
                                >{\centering\arraybackslash}X 
                                >{\centering\arraybackslash}X}
            \toprule
            No. & ${{\varepsilon }_{r}}$ (mm) & ${{\varepsilon }_{a}}$ (mm) \\ 
            \toprule

            Case 1 & +0.011 & +0.003 \\

            Case 2 & $-0.026$ & +0.009 \\

            Case 3 & $-0.013$ & +0.007 \\
   
            \bottomrule
        \end{tabularx}
    \end{table}

        \subsubsection{Predictive performance}
            \label{subsubsec_case_2_Predictive}

            \begin{figure}[ht]%
                \centering
                \includegraphics[width=0.99\textwidth]{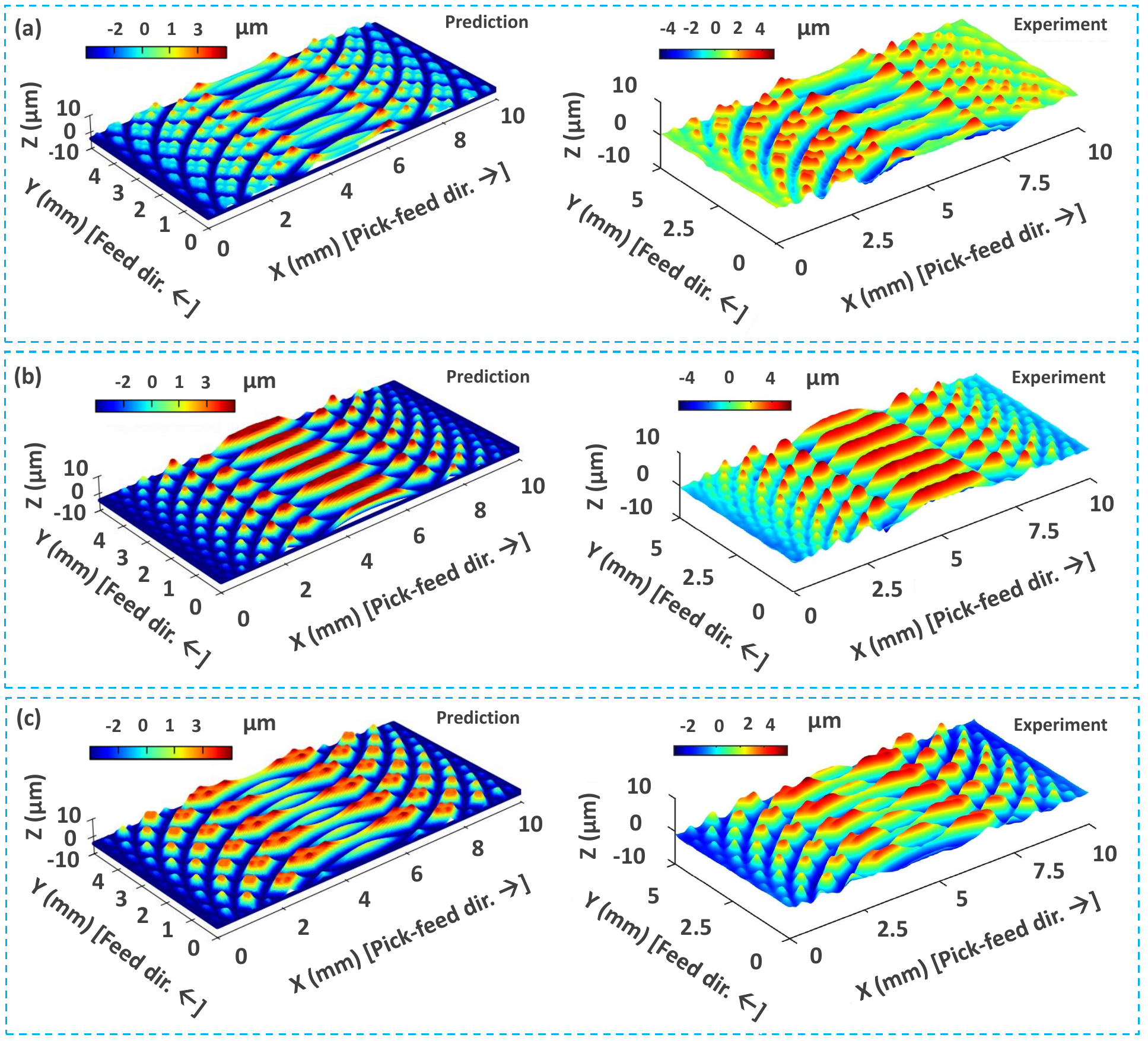}
                \caption{3D surface topographies of predictions and experiments for (a) Case 1, (b) Case 2, and (c) Case 3. Experimental data are adapted from \cite{wang2023high}.}\label{3D_Surface_Topography}
            \end{figure}

            The EFSM predictions are evaluated under three cutting conditions listed in \autoref{tab_milling_paramters} and compared with the corresponding experimental results, as shown in \autoref{3D_Surface_Topography}. For the different cases, the predictions show good agreement with the experimental results in terms of both the generated surface patterns and the corresponding surface contour values.
            
\begin{figure}[ht]%
                \centering
                \includegraphics[width=0.99\textwidth]{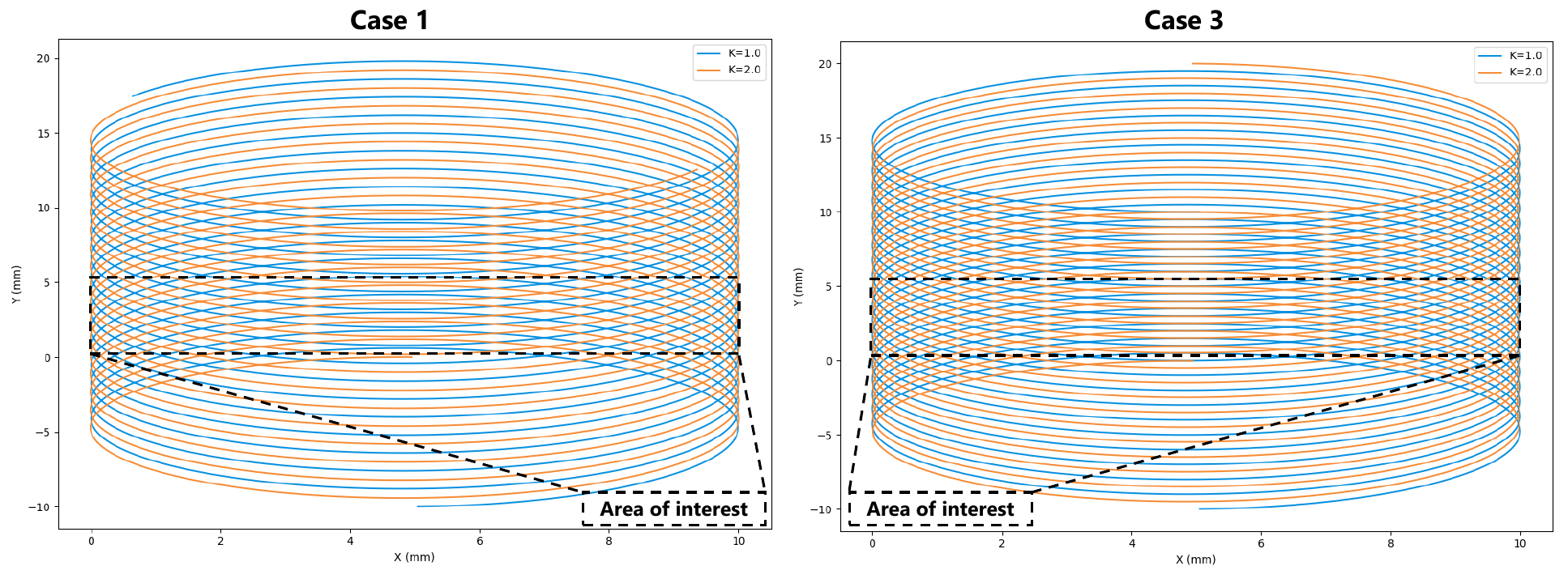}
                \caption{Trajectory maps are generated by the cutting-edges for the various cases. The area of interest corresponds to the region from which the surface topographies were captured. $K$ denotes the index of the cutting-edge.}\label{Trajectories}
            \end{figure}

            The peaks and valleys observed on the machined surface are mainly caused by the front-cutting and back-cutting motions of the tool. These specific patterns can be related to the cutting-edge trajectories, as illustrated in \autoref{Trajectories}. Each cutting-edge follows a unique trajectory, and the combined effect of tool rotation and feed motion results in the formation of peaks and valleys on the surface. Considering the trajectory maps for the different cases, similar surface topography patterns are generated. In regions where cutting-edge trajectories are sparse or absent, higher peaks tend to dominate, whereas areas with dense trajectory intersections generally exhibit lower surface heights.

\begin{figure}[ht]%
                \centering
                \includegraphics[width=0.99\textwidth]{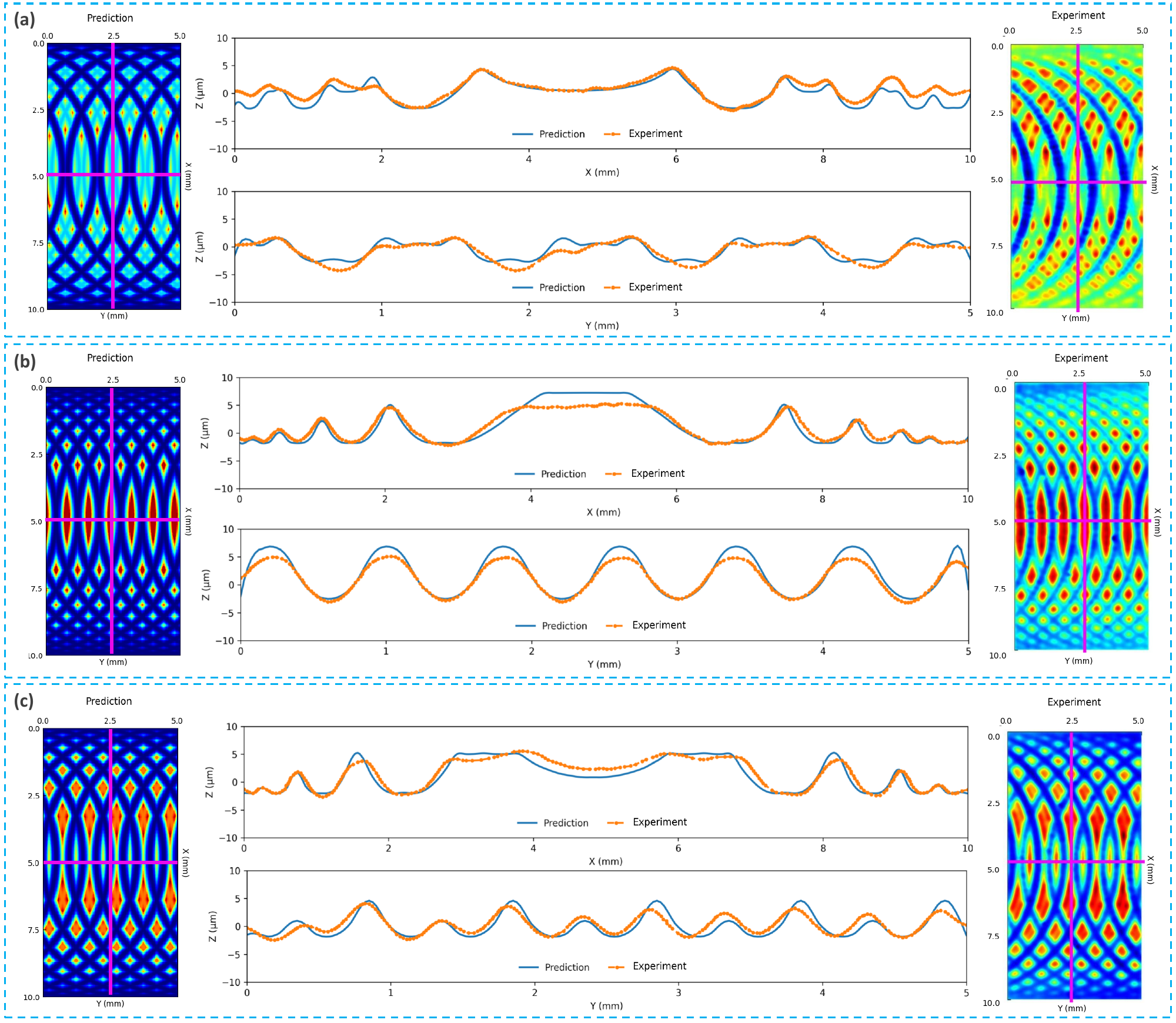}
                \caption{2D roughness maps obtained from the top view of the 3D surface topography, together with the corresponding 2D contour lines along the feed and pick-feed directions for (a) Case 1, (b) Case 2, and (c) Case 3. Experimental data are adapted from \cite{wang2023high}.}\label{2D_Contour_Lines}
            \end{figure}

            To provide a better understanding of the generated patterns, \autoref{2D_Contour_Lines} illustrates the 2D contour lines extracted along the feed and pick-feed directions from the 3D topography images for all case studies. It can be observed that periodic behavior appears along the feed direction; however, this pattern is absent in the pick-feed direction, where a gradual variation is observed instead. The roughness contours exhibit comparable behaviors, such as the presence of peaks and valleys, with only minor discrepancies in certain regions. \autoref{2D_Contour_Lines} also highlights the similarity between the predicted and experimental results from a top-view 2D representation.

\begin{figure}[ht]%
                \centering
                \includegraphics[width=0.99\textwidth]{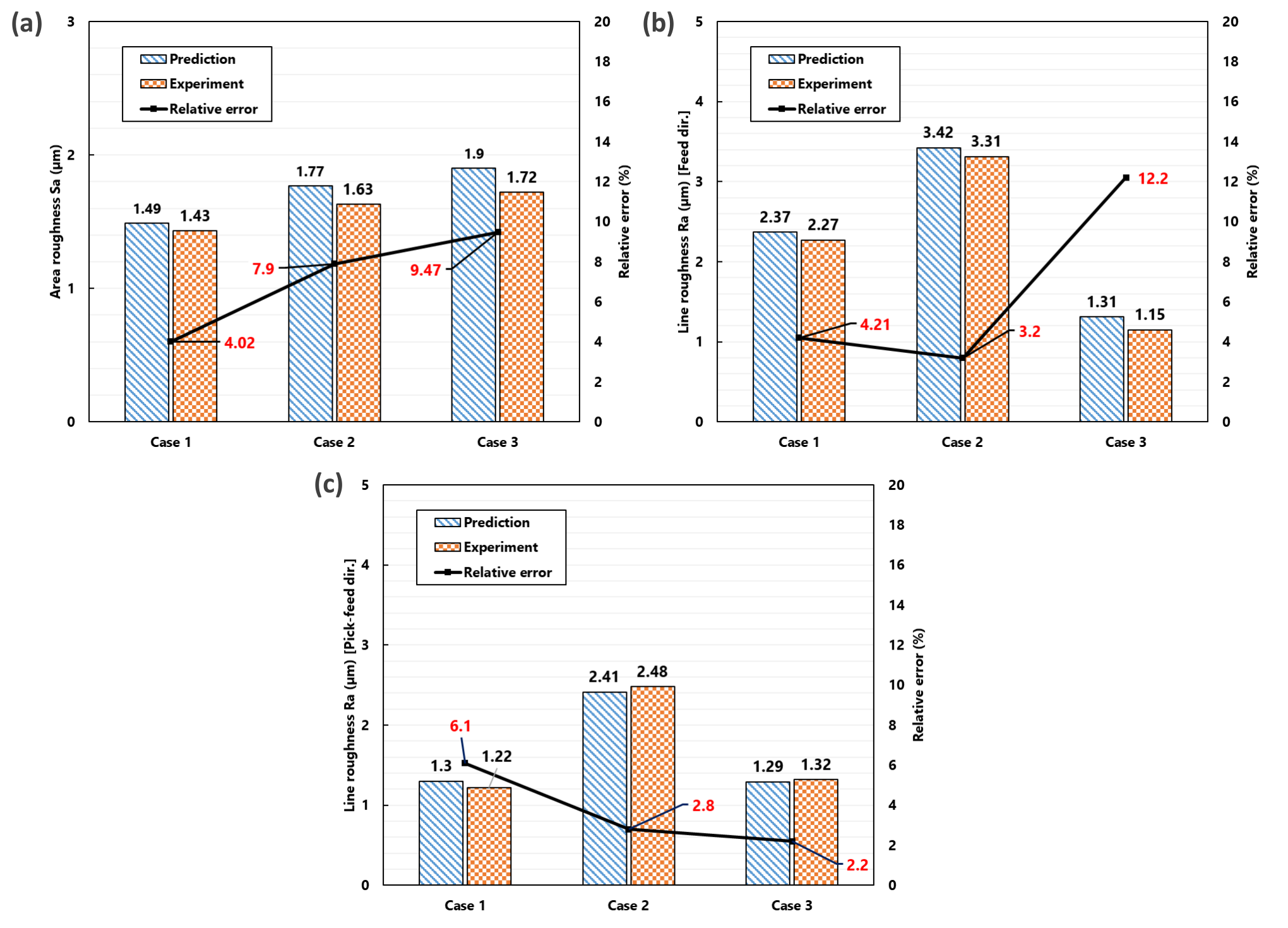}
                \caption{Comparison of predicted and experimental values for (a) area roughness and the corresponding relative error, (b) line roughness and relative error along the feed direction, and (c) line roughness and relative error along the pick-feed direction. Experimental data are adapted from \cite{wang2023high}.}\label{Roughness_Bar_Chart}
            \end{figure}

            To further quantify the accuracy of the model, the average surface roughness ($S_a$) and line roughness ($R_a$) along the feed and pick-feed directions are compared with the corresponding experimental measurements. As shown in \autoref{Roughness_Bar_Chart}, the predicted and experimental average values of area roughness and line roughness exhibit good agreement for all investigated cases. The relative errors for most cases remain low, with a maximum value of approximately 12$\%$ and average value of 5.7$\%$, indicating the strong predictive performance of the proposed model.

    \subsection{Computational performance}
        \label{subsec_computational_performance}

        The computational performance is characterized based on the execution time required to simulate identical milling conditions under the same computer configuration and hardware environment. The following computational cost evaluation was conducted on a laptop with the following specifications: operating system: Windows 11 Pro; software: open-source programming languages including C++ and Python; CPU: 13th Gen Intel(R) Core(TM) i5-1335U (1.30 GHz); GPU: Intel® Iris® Xe Graphics; and RAM: 16 GB.
        
        The surface area of interest considered in the simulations is $10\times 5\,m{{m}^{2}}$, consistent with the case 2. For consistency, the analysis assumes zero tool run-out. As highlighted in \autoref{tab_milling_paramters}, three distinct values are defined for each cutting parameter. By forming combinations of these values, nine unique cutting conditions are obtained, which are presented in \autoref{tab_selected_milling_paramters}. The computational performance of the conventional FSM and the proposed EFSM models is evaluated based on these cutting conditions.

        \begin{table}[h]
            \centering
            \fontsize{8}{13}\selectfont
            \caption{Selected milling process parameters for computational efficiency evaluation.}
            \label{tab_selected_milling_paramters}
            % \begin{tabularx}{0.5\textwidth}{X X X X}
            \begin{tabularx}{0.5\textwidth}{>{\centering\arraybackslash}X 
                                >{\centering\arraybackslash}X 
                                >{\centering\arraybackslash}X 
                                >{\centering\arraybackslash}X}
                \toprule
                No. & ${{v}_{c}}$ (m/min) & ${{f}_{z}}$ (mm/tooth) & ${{a}_{P}}$ (mm) \\ \toprule
    
                 1 & 170 & 0.4 & 0.3 \\
    
                 2 & 170 & 0.5 & 0.4 \\
    
                 3 & 170 & 0.6 & 0.5 \\

                 4 & 200 & 0.4 & 0.3 \\

                 5 & 200 & 0.5 & 0.4 \\

                 6 & 200 & 0.6 & 0.5 \\

                 7 & 230 & 0.4 & 0.3 \\

                 8 & 230 & 0.5 & 0.4 \\

                 9 & 230 & 0.6 & 0.5 \\
       
                \bottomrule
            \end{tabularx}
        \end{table}

        \autoref{tab_computational_performance} highlights the number of trajectory points for the selected case studies and the corresponding computation time for both the FSM and EFSM models. It should be noted that the reported computation time only accounts for the execution time of the main simulation process used to obtain the surface topography and does not include the entire workflow, which also involves post-processing tasks. The number of trajectory points represents the model size, while the computation time reflects the computational efficiency of the models. Each row in \autoref{tab_computational_performance} corresponds to a test case in which the same number of trajectory points is generated for both FSM and EFSM models, along with their corresponding computation times. As can be observed, an improvement ranging from approximately 40x to 50x is achieved using the EFSM, demonstrating the high computational efficiency of the proposed model. Across all case studies, an average speedup factor of 42.2x in computation time is obtained.

        To further investigate the efficiency of the proposed model, the same milling conditions were simulated using 10x more trajectory points, resulting in higher-resolution surface topographies. \autoref{tab_computational_performance} also lists the speedup factors when a greater number of trajectory points are considered. From the table, it can be seen that the speedup factor increases in some cases by nearly an order of magnitude, indicating even better computational performance. This behavior is related to the Python implementation, which exhibits a disproportionate increase in computational time as the number of trajectory points grows. This is mainly attributed to interpreter overhead, dynamic memory management, and reduced cache efficiency in Python, whereas the C++ implementation benefits from compiled execution and more efficient memory handling.

        \begin{table}[h]
            \centering
            \fontsize{8}{13}\selectfont
            \caption{Number of trajectory points and computational performance, measured by execution time, for FSM and EFSM under various process conditions.}
            \label{tab_computational_performance}
            
            \begin{tabularx}{\textwidth}{
            >{\centering\arraybackslash}p{1.2cm}
            >{\centering\arraybackslash}p{3.2cm}
            >{\centering\arraybackslash}p{2cm}
            >{\centering\arraybackslash}p{2cm}
            >{\centering\arraybackslash}p{1.5cm}
            >{\centering\arraybackslash}p{3.2cm}}
        
                \toprule
                \multirow{2}{*}{No.} & \multirow{2}{*}{Number of trajectory points} & \multicolumn{2}{c}{Computation time (s)} & \multirow{2}{*}{Speedup (x)} &  \multirow{2}{*}{Speedup (10× trajectory points)} \\ 
                \cmidrule(lr){3-4}
                 &  & FSM & EFSM \\ 
                \midrule
                
                1 & 693000 & 29.96 & 0.81 & 36.9 & 50.8 \\
                2 & 554400 & 29.67 & 0.68 & 43.6 & 54 \\
                3 & 462000 & 23.37 & 0.55 & 42.5 & 56.4 \\
                4 & 589050 & 27.57 & 0.65 & 42.4 & 51.1 \\
                5 & 471240 & 22.36 & 0.52 & 43   & 50 \\
                6 & 392700 & 22.69 & 0.45 & 50.4 & 50.2 \\
                7 & 512218 & 23.86 & 0.59 & 40.4 & 47.3 \\
                8 & 409774 & 19.27 & 0.46 & 41.8 & 52.6 \\
                9 & 341478 & 16.87 & 0.43 & 39.2 & 49.2 \\
            
            \bottomrule
            \end{tabularx}
        \end{table}

        One important application of improving computational efficiency is to accelerate the dataset generation process, which can be used for data augmentation in machine learning and data-driven analysis. The main idea is to consider all influential parameters in the machining process, such as tool geometry and cutting parameters, and define a plausible range for each parameter. A large number of datasets can then be generated from all possible parameter combinations, typically using random sampling strategies such as Latin Hypercube Sampling (LHS).
        
        The EFSM is therefore also developed to efficiently generate large datasets consisting of 3D surface topography images or 2D surface roughness maps, together with their corresponding coordinate systems and input machining conditions. It is adaptable to various tool types and milling parameters, enabling the model to cover a wide range of parameter combinations by simply incorporating the kinematic equations of the specific tool into the formulation.

%---------------------------------------------------------------------------------------------
%---------------------------------------------------------------------------------------------
%---------------------------------------------------------------------------------------------

\section{Conclusions}
    \label{conclusions}

    In this work, an optimized open-source framework is developed to efficiently and accurately capture 3D surface topography and roughness parameters in milling operations. Leveraging the EFSM approach, the framework focuses on algorithmic optimization to achieve high computational performance, making it suitable for large-scale surface simulations and the generation of extensive datasets for data-driven modeling. The following conclusions can be drawn:

    \begin{enumerate}

    \item The EFSM framework significantly enhances computational performance in milling simulations by optimizing algorithms while maintaining high modeling accuracy.

    \item The hybrid C++/Python architecture executes computationally intensive operations in compiled C++ to minimize interpreter overhead, while Python provides flexibility for configuration, control, and post-processing.

    \item The modular design including tool geometry, kinematics, surface computation, and trajectory modules ensures both clarity and efficiency. Preallocated memory and cache-friendly data structures further reduce runtime overhead.

    \item Stable runtime behavior is achieved by avoiding dynamic memory allocation in inner computational loops and leveraging precomputed constants and efficient matrix operations.

    \item The open-source implementation provides a reusable and extensible platform for large-scale surface simulation and dataset generation, supporting the development of data-driven surrogate models across diverse milling operations.

    \end{enumerate}

%---------------------------------------------------------------------------------------------
%---------------------------------------------------------------------------------------------
%---------------------------------------------------------------------------------------------

\section*{Declaration of competing interest}
	
   The authors declare that they have no known competing financial interests or personal relationships that could have appeared to influence the work reported in this paper.

%---------------------------------------------------------------------------------------------
%---------------------------------------------------------------------------------------------
%---------------------------------------------------------------------------------------------
\section*{Acknowledgments}

    The research outcomes presented herein were achieved within the framework of the SuPreAM project, supported by funding from the European Union's Research Fund for Coal and Steel (RFCS), project number 101112346.

\section*{Data availability}
    The datasets generated and analyzed during this study are available from the corresponding author upon reasonable request.

\section*{Code availability}

The code for this paper is available on GitHub \url{https://github.com/HadiBakhshan/surf-topo.git}.

%---------------------------------------------------------------------------------------------
%---------------------------------------------------------------------------------------------
%---------------------------------------------------------------------------------------------

	\bibliographystyle{elsarticle-num} 
	\bibliography{Bibliography}
	
\end{document}